\documentclass[%
superscriptaddress,
 amsmath,amssymb,
prl,
twocolumn,
]{revtex4-2}
\usepackage[export]{adjustbox}
\usepackage{graphicx}
\usepackage{color}
\usepackage{xcolor}
\usepackage{natbib}
\usepackage{wasysym}
\usepackage{dcolumn}
\usepackage{bm}
\usepackage{hyperref}
\definecolor{mygreen}{rgb}{0,0.5,0}
\definecolor{mybrown}{rgb}{0.65,0.16,0.16}

\def\beq {\begin{equation}}
\def\eeq {\end{equation}}
\def\beqa {\begin{eqnarray}}
\def\eeqa {\end{eqnarray}}
\def \bnum {\begin{enumerate}}
\def \enum {\end{enumerate}}
\def\bi {\begin{itemize}}
\def\ei {\end{itemize}}

\def\rel {R_{\lambda}}

\def\la {\langle}
\def\ra {\rangle}

\def\modsp{|\la (\delta_r u )^n \ra|^{1/n}}
\def\spmod{\la |\delta_r u |^n \ra^{1/n}}
\def\normr{r/\eta}
\def\dru{(\delta_r u)^n}
\def\drumod{|\delta_r u|^n}
\def\drup{(\delta_r u^{+})^n}
\def\drum{(\delta_r u^{-})^n}
\def\dur{\delta_r u}
\def\durm{|\delta_r u|}
\def\dup{\delta_{r} u^+}
\def\dum{\delta_{r} u^-}

\begin{document}
\title{The asymmetry of velocity increments in turbulence}

\author {Katepalli R. Sreenivasan} 
\affiliation {Department of Mechanical and Aerospace Engineering, New York University, New York, NY, $11201$, USA}
\affiliation {Department of Physics and the Courant Institute of Mathematical Sciences, New York University, New York, NY $11201$, USA}
\email{krs3@nyu.edu}

\author {Kartik P. Iyer} 
\affiliation {Department of Physics, Michigan Technological University, Houghton, MI $49931$}
\affiliation {Department of Mechanical Engineering-Engineering Mechanics, Michigan Technological University, Houghton, MI $49931$}
\email{kiyer@mtu.edu}

\author {Ashvin Vinodh} 
\affiliation {Department of Mechanical Engineering-Engineering Mechanics, Michigan Technological University, Houghton, MI $49931$}

\begin{abstract}
We use well-resolved direct numerical simulations of high-Reynolds-number turbulence to study a fundamental statistical property of turbulence --- the asymmetry of velocity increments --- with likely implications on important dynamics. This property, ignored by existing small-scale phenomenological models, manifests most prominently in the non-monotonic trend of velocity increment moments (or structure functions) with the moment order, and in differences between ordinary and absolute structure functions for a given separation distance. We show that high-order structure functions arise nearly entirely from the negative side of the probability density of velocity increments, essentially removing the ambiguity between ordinary and absolute moments, and provide a plausible dynamical interpretation of this result. 

\end{abstract}
\maketitle
\noindent
Velocity fluctuations in three-dimensional homogeneous and isotropic turbulence have a symmetric distribution but it is known since Kolmogorov's seminal work \cite{K41b} that the longitudinal velocity differences are asymmetric. This characteristic imposes a structure on turbulence, which is often interpreted as an important indicator of vortex stretching and a forward cascade of energy. But what are its additional consequences? This paper addresses the question and its fundamental significance. We answer it particularly in the context of structure functions, which are moments of velocity increments over a chosen separation distance. We use very large and well-resolved direct numerical simulations of high-Reynolds-number turbulence for the purpose. 

\vspace{0.15cm}
\paragraph{Background:} Structure functions of turbulence, as originally introduced by Kolmogorov \cite{K41a}, are ordinary moments of velocity increments $\delta_r u$ ($u$ being a velocity component) over two space points chosen with the separation distance $r$. Numerous other authors since then (some will be mentioned below) have considered only absolute moments, while implying that they have the same scaling properties as ordinary moments. That is, if
\beq
\la \dru \ra  = C_1  (r/L)^{\zeta_n} \;,\ 
\la \drumod \ra = C_2  (r/L)^{\xi_n} \;,
\eeq
where $C_1$ and $C_2$ are $r$-independent prefactors, 
the presumption has been that $\zeta_n = \xi_n$ for $n \ge 0$. 

\vspace{0.15cm}
Thus, most phenomenological models \cite{k62,Benzi1984,CMKRS87,SL94,Yakhot01,Ruelle12,oz15,lb19} consider only absolute moments. An influential example is the so-called Extended Self-Similarity \cite{Benzi1993}, which plots structure functions of an arbitrary order against the absolute third moment, with the assumption that the latter will scale as the ordinary third order structure function --- which varies linearly with an inertial separation \cite{K41b}. This assumption is only roughly true \cite{krs98,dhruva2000,gotoh2002}. The difference for the so-called longitudinal structure functions (i.e., the velocity $u$ in the direction of the separation distance $r$) arises because the probability density function (PDF) of $\delta_r u$ is not symmetric for inertial range separation (also in the dissipative range \cite{MY75}), and one does not know {\it a priori} whether absolute moments, which arise from sums of positive and negative sides of the PDF of $\delta_r u$, would scale the same way as ordinary moments, which are differences between those contributions. 

\vspace{0.15cm}
\paragraph*{Data:} In this work, we use from direct numerical simulations (DNS) of incompressible Navier-Stokes equations in a periodic cube of length $L_0 = 2\pi$ using $N^3$ collocation points. The simulations are forced to a statistically steady state, which was attained by forcing the low wavenumber magnitudes $k \le 2$ using the scheme described in Ref.~\cite{DAD2010}. At least ten snapshots of the DNS saved in the statistically steady state, over more than eight eddy-turn over times $L/u^\prime$ where $L \approx 0.2 L_0$ is the integral scale and $u^\prime$ is the root-mean-square velocity fluctuation, were used for the analysis. The 
Reynolds number $\rel = u^\prime \lambda/\nu$ is given in terms of the Taylor-microscale $\lambda$ \cite{krs83} and the kinematic viscosity $\nu$. Averages were performed over space, time and the solid angle, as in \cite{KIKRS20}. More details on the DNS data, including temporal and spatial resolution checks, as well as convergence of statistical averages, can be found in Ref.~\cite{KIKRS20}. We show results mainly from the $4096^3$ DNS at $\rel = 650$ with small-scale resolution $\Delta/\eta \approx 1.1$ where $\Delta$ is the uniform grid-spacing and $\eta = (\nu^3/\epsilon)^{1/4}$ is the Kolmogorov length scale, where $\epsilon$ is the average dissipation rate. Our conclusions hold for the $8192^3$ data as well (but the latter are more sparsely processed because of computing time requirements). Despite these and other confidence-building checks, the precise numerical values of very high-order quantities may have uncertainties which we cannot completely quantify, but our qualitative conclusions have withstood repeated scrutiny.   
 
\begin{figure}
\includegraphics[width=0.5\textwidth,center]{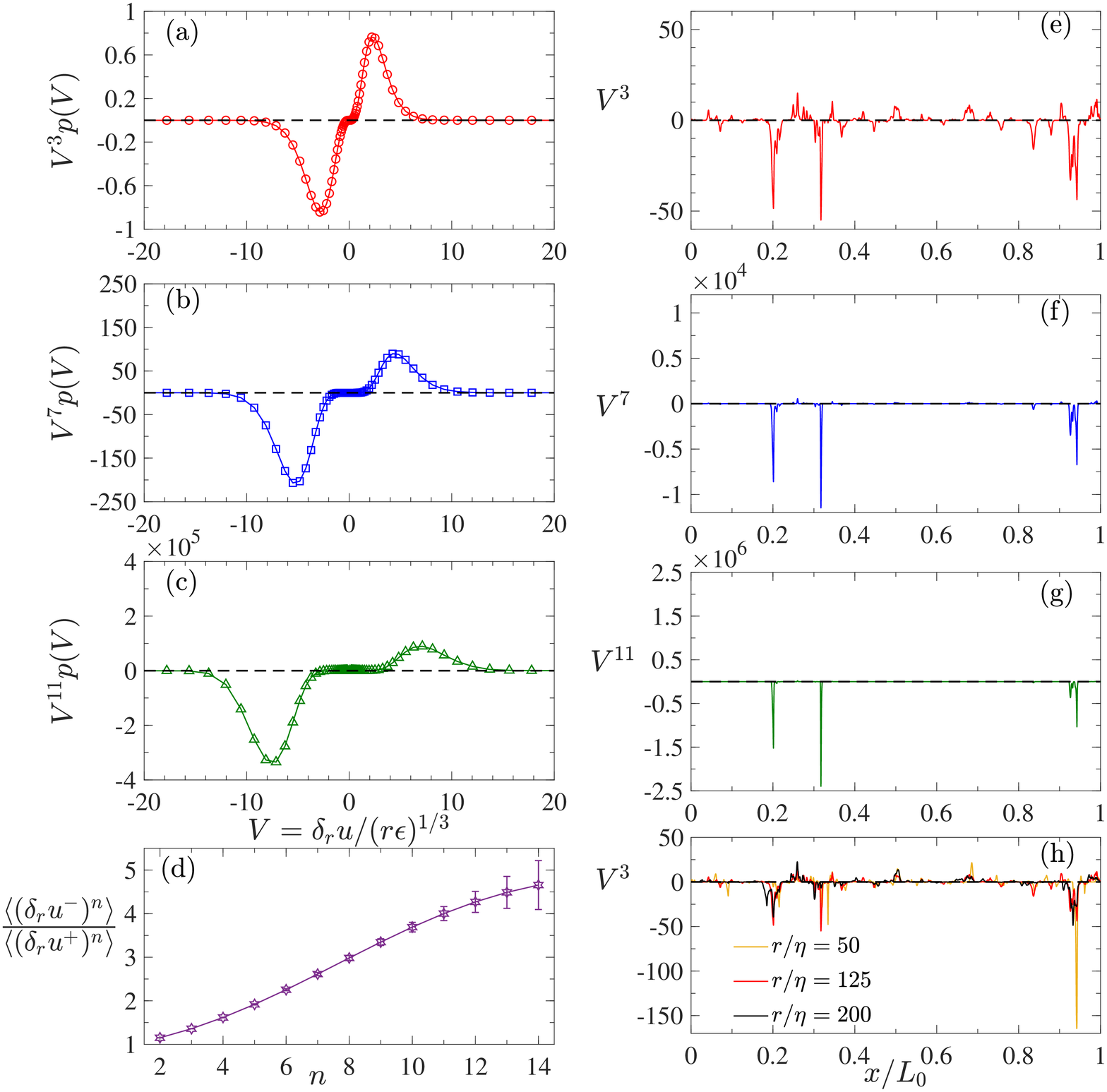}
\protect\caption{Integrands of odd-order moments of the non-dimensional longitudinal increment $V = \delta_r u/ (r \epsilon )^{1/3}$ for orders $3$, $7$ and $11$ are shown in panels (a), (b) and (c), respectively, for inertial range separation $r/\eta = 125$ from $4096^3$ DNS at $\rel = 650$ in a cube of edge-length $L_0$. The average energy dissipation rate $\epsilon$ equals the interscale energy transfer rate. Panel (d) is the ratio of contributions from the negative and positive parts of the PDF of $\delta_r u$, with increasing moment order. Panels (e), (f) and (g) are instantaneous line traces along the $x$-direction, contributing to moments $\la V^3 \ra$, $\la V^7 \ra$ and $\la V^{11} \ra$, respectively. The reference line at zero is given in all panels. Panel (h) shows the energy flux at three different scales in the inertial range, demonstrating that the large excursions are located around the same spatial positions at all inertial-range scales. This suggests the presence of coherent structures responsible for the large amplitudes of forward energy cascade.}
\label{f1.fig}
\end{figure}

\begin{figure}
\includegraphics[width=0.45\textwidth,center]{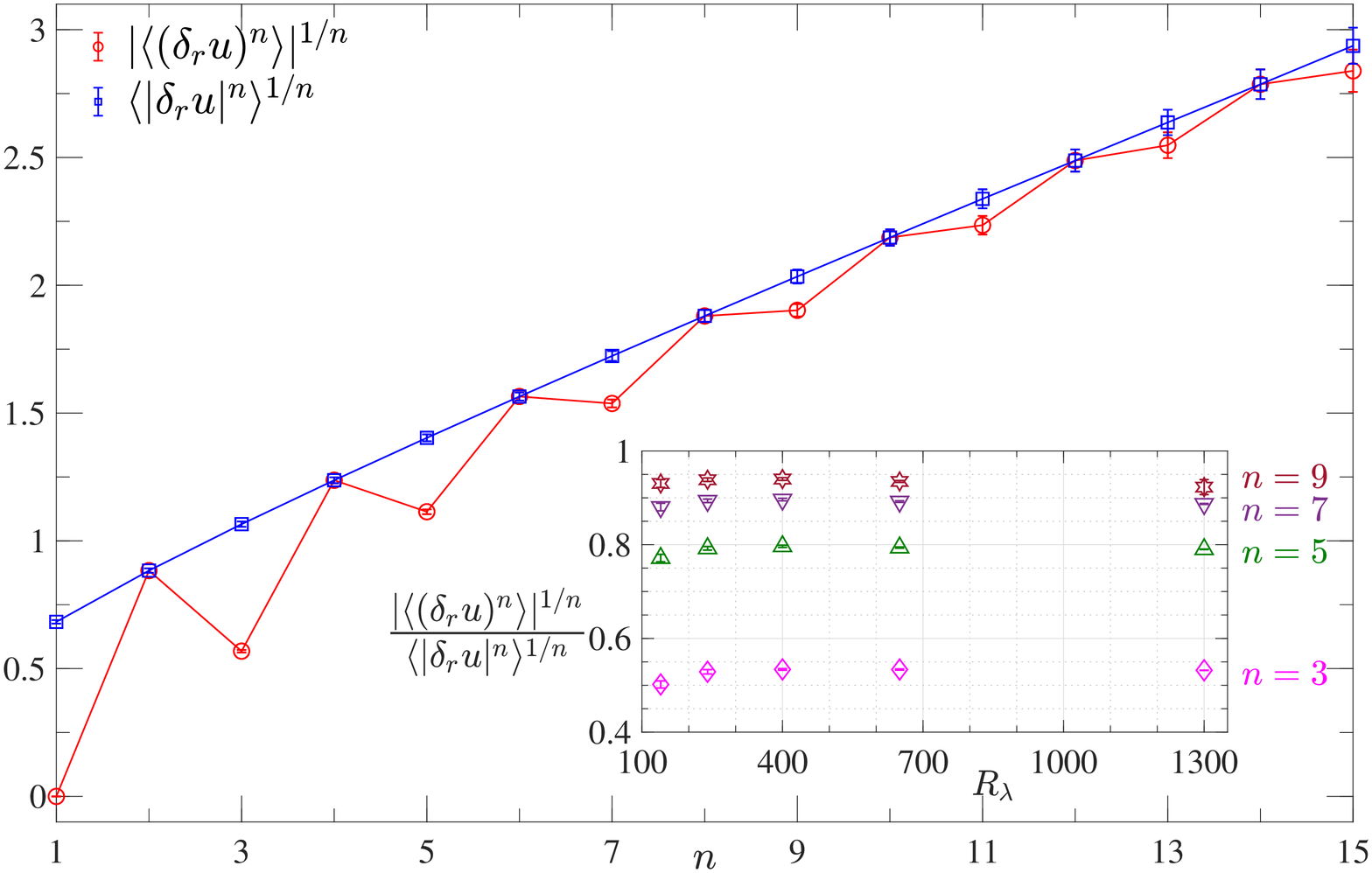}
\protect\caption{Longitudinal structure functions $\modsp$ and $\spmod$ as functions of order $n$ for a $4096^3$ cube, $\rel = 650$, with the scale-separation fixed at $\normr \approx 125$. Inset shows their ratio
for odd-orders $n=3,5,7$ and $9$ as functions of the Taylor microscale Reynolds number $\rel$.
Error bars (too small to stand out) are the standard deviation of temporal variations of the moments in the statistically steady state.}
\label{f2.fig}
\end{figure}
\vspace{0.15cm}
\paragraph{Asymmetry of structure functions:} To study the asymmetry of $\dur$ we can decompose it into positive and negative parts \cite{KRS94} as
\begin{equation}
\label{dup.eq}
\dup =  \frac{1}{2}[|\delta_r u|+\delta_r u]~; \;\;\;
\dum  =  \frac{1}{2}[|\delta_r u|-\delta_r u].    
\end{equation}
Both $\dup$ and $\dum$ are non-negative, represent the magnitudes of the
positive and negative parts of $\dur$, respectively, and follow the relation
\beq
\label{stfn.eq}
\la \dru \ra  =  \la \drup \ra + (-1)^n \la \drum \ra \;,
\eeq 
for $n \ge 1$. It has been shown elsewhere \cite{krs96} that
the positive and negative increment contributions also display robust inertial range power laws, with exponents $\zeta_n^\pm$ and the respective power-law prefactors $C_3$ and $C_4$, as
\beq
\la \drup \ra = C_3  (r/L)^{\zeta_n^+} \;,\
\la (\delta_r u^-)^n \ra = C_4 (r/L)^{\zeta_n^-} \;.
\eeq
Since $\dup$, $\dum$ and $\durm$ are all non-negative, application of H\"older inequality for high $\rel$ or $r/L \ll 1$ shows that the corresponding scaling exponents $\zeta_n^\pm$ and $\xi_n$ are concave functions of the moment order $n$ \cite{Fri95}. However, concavity does not hold for the longitudinal exponents $\zeta_n$ with respect to $n$, since the sign of the longitudinal moments $\la \dru \ra$ depends on whether $n$ is even or odd. 
 
\begin{figure}
\includegraphics[width=0.45\textwidth,center]{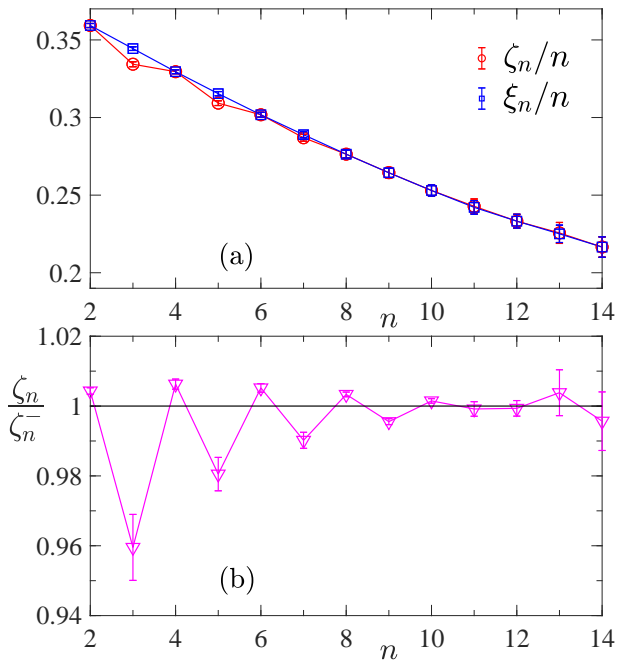}
\protect\caption{(a) Scaling exponents $\zeta_n/n$ and $\xi_n/n$ for ordinary and absolute structure functions, as a function of order $n$, for the longitudinal case. As before, the error bars, too small to stand out, represent the standard deviation of local slopes in inertial-range power laws.
(b) Ratio of the scaling exponents $\zeta_n$ of the longitudinal structure function and $\zeta_n^-$ which is the corresponding exponent of its negative part, for varying moment order, $n$. Horizontal line at unity is given for reference. Error bars represent the standard deviation of the ratio of local slopes of the respective inertial range power laws.}
\label{f3.fig}
\end{figure}

\vspace{0.15cm}
The imbalance between the $\dup$ and $\dum$ contributions to the moments $\la \dru \ra$ is illustrated in Fig.~\ref{f1.fig}. The left panels show the integrands for three odd moments, $n = 3, 7$ and $11$, for the quantity $\dur/(r\epsilon)^{1/3}$, corresponding to $r/\eta = 125$ (which lies within the inertial range) at $\rel = 650$. The figures show the increasing dominance of the contribution of $\dum$ over the that of $\dup$, as the moment order increases. (The situation is the same for even moments as well, but there are no cancellations between negative and positive parts.) In fact, panel (d) shows that how ratio of the magnitude of the negative part of the PDF to its positive part increases monotonically with the order of the moment. It is not certain if this ratio increases without bound or saturates for moments beyond a certain order because of resolution and sampling uncertainties at large orders, but the increasing irrelevance of the positive part of the PDF with increasing $n$ is incontrovertible.

\vspace{0.15cm}
\paragraph{Intermittency in energy transfer rates:} Another comment relates to the intermittent nature of energy transfer, shown in panel (e) of Fig.~\ref{f1.fig}; the net energy transfer across scales is not a simple unidirectional process but a small difference (notice the scale on the ordinate) between two highly unbalanced backward and forward transfer processes; see \cite{krs1999,sualeh}.

\vspace{0.15cm}
The right panels of Fig.~\ref{f1.fig} show the one-dimensional cuts of the third, seventh and the eleventh power of $\dur/(r\epsilon)^{1/3}$. For large $n$, it appears clear that the contributions to the moments of $\delta_r u$ come essentially from its negative parts. These traces show the intermittent feature that is not obvious from the left panels: $(\delta_r u)^3/r$, which represents the local energy transport in real space, is a highly intermittent process. Another trace taken at another time reveals the same characteristics but at different spatial locations. If one imagines a large shower head with many pinholes that allow water flow, only some pinholes are open at any given time, with different set of pinholes open at the next moment. 

\vspace{0.15cm}
The intermittency and cancellation for odd-order moments results in a non-monotonic, zig-zag pattern in the moments of $|\la \dru \ra|^{1/n}$ as shown in Fig.~\ref{f2.fig} with smaller values for odd moments compared to even-orders to either side (which are free of cancellations), as has been seen before \cite{vdw95,dhruva2000}. 
The ratio of the ordinary to the absolute moments, bounded above by unity, show at most a weak dependence on $\rel$ at a fixed odd value of $n$, as seen in the inset of Fig.~\ref{f2.fig}. This result indicates that this zig-zag pattern is a robust feature of the longitudinal increment field.
This pattern diminishes at higher orders (though it does not vanish) as seen in the main part of Fig.~\ref{f2.fig}, because the degree of cancellation between negative and positive parts decreases (Fig.~\ref{f1.fig} (d)). 
There is a clear suggestion from the inset of Fig.~\ref{f2.fig} that the zig-zag behavior is independent of the Reynolds number for high Reynolds numbers.

\vspace{0.15cm}
The question is whether the zig-zag nature of Fig.~\ref{f2.fig} carries over to inertial-range scaling exponents as well. Figure~\ref{f3.fig} (a) plots $\zeta_n/n$ as a function of order $n$. For low-orders the same saw-toothed behavior of Fig.~\ref{f2.fig} exists in the corresponding scaling exponents. It says that the longitudinal scaling exponents $\zeta_n$ violate the concavity property with respect to the moment order, locally at low orders. However, this feature diminishes for large $n$, so there is ultimately no ambiguity with respect to the scaling exponents when the moment order is large.  

\vspace{0.15cm}
Since the high-order moments (both even and odd) mainly stem from the negative part $\dum$, as already shown in Fig.~\ref{f1.fig} (a)-(d), it follows from Eq.~\ref{stfn.eq} that,
\beq
\label{highordapprox.eq}
\la \dru \ra  \approx (-1)^n \la \drum \ra \;,
\eeq 
for large $n$.

\vspace{0.15cm}
Not only do the contributions of $\dum$ approximate the undecomposed quantity $\dur$ at high orders but their respective exponents $\zeta_n$ and $\zeta_n^-$ also match, to an excellent degree, for orders $n \ge 8$ as shown in Fig.~\ref{f3.fig}(b). Since $\zeta_n$ are expected to be $\rel$-independent at sufficiently large $\rel$, $\zeta_n^-$ for large-$n$ (say $n \gtrsim 8$) also share this same property. Importantly, since $\zeta_n^-$ are concave in $n$ it follows that the longitudinal exponents $\zeta_n$  are also concave functions in $n$ for $n > 8$, say, to a good approximation. We note in passing that, although the contribution of the positive increment $\dup$ diminishes with increasing order (see Fig.~\ref{f1.fig}(d)), the exponents $\zeta_n^+$ are slightly larger than $\zeta_n^-$, thus closer to the self-similar value of $n/3$ (though they remain distinctly smaller \cite{krs96}).

\begin{figure}
\includegraphics[width=0.5\textwidth,center]{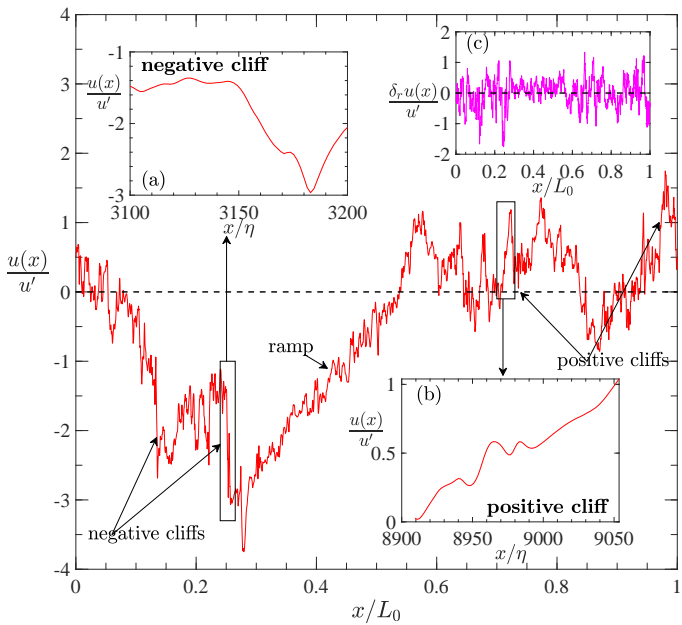}
\protect\caption
{A typical instantaneous trace of the $x$-component velocity along the $x$-direction normalized by the root-mean-square velocity as a function of $x/L_0$, where $L_0$ is the box length (with $8192$ grid points to a side), and $\rel=1300$. The trace is dominated by steep negative cliffs
(which contribute to $\dum$) and gentler positive cliffs (which contribute to 
$\dup)$ connected by ramp-like structures. Inset (a) expands the region $x/L_0 \in (0.2,0.3)$ (marked by the rectangle) in terms of 
$x/\eta$ to reveal the negative-cliff structure while inset (b) expands $x/L_0 \in (0.70,0.72)$ (marked by the rectangle) in terms of 
$x/\eta$ to highlight the positive-cliff structure. Negative cliffs are consistently much deeper than the positive ones. 
Inset (c) shows the normalized velocity difference $\delta_r u = u(x+r)-u(x)$ for $r/\eta = 175$ as a function of $x/L_0$. 
Horizontal dashed lines at zero is given for reference.
}
\label{f4.fig}
\end{figure}

\vspace{0.15cm}
\paragraph{Velocity signals:} The pronounced negative asymmetry at large orders, clearly seen as huge negative excursions in longitudinal increments, must be detectable also in the velocity signal. It is clearly seen in the instantaneous (but representative) velocity trace in Fig.~\ref{f4.fig} that negative cliffs contributing to negative increments are connected to positive cliffs that contribute to positive increments by ramp-like structures. A expanded view of one of the negative cliffs in inset (a) of Fig.~\ref{f4.fig} shows that these dominant structures resemble
the Khokhlov saw-tooth solution of the one-dimensional Burgers equation \cite{khokhlov61}. In comparison, a magnification of positive cliffs (see inset (b)) shows that the positive cliffs are less sharp and tend to differ in structure from negative cliffs.
The corresponding trace of velocity increment, shown in inset (c), verifies that the ramps are, as expected, characterized by depleted velocity differences while the cliffs (especially the negative ones) correspond to large gradients. 

\vspace{0.15cm}
We have demonstrated here that the anti-alignment property of velocity increments with their separation vectors favors larger
negative velocity jumps $\dum/u^\prime \gg 1$ than positive jumps. Such an alignment scheme  
imposes a preferential structure on the velocity field with large negative cliffs and gentler positive cliffs interspersed with ramp-like structures. This means that for large orders or, equivalently for large velocity increments $|\delta_r u|/u^\prime \gg 1$,
the negative increment field $\dum$ approximates the velocity difference field to a reasonable approximation. 
As already noted, these small-scale negative velocity jumps resemble the situation for Burgers turbulence with shocks \cite{bec}. Similar structures are also seen in passive scalar fields advected by turbulence \cite{celani,kikrs18}, and in transverse velocity increments \cite{staicu,KIKRS20}.  
The approximation of the velocity difference field by the negative increment field at large-orders also has important theoretical consequences \cite{krsvy,TD22}.

\vspace{0.15cm}
A (somewhat speculative) point of this work is that the net energy transfer across scales is the slight imbalance between two oppositely directed fluxes. The persistence of large-amplitudes fluxes across the entire inertial range, as shown in Fig.~\ref{f1.fig}(h), suggests a connection to spatially coherent activity, likely the ubiquitously observed vortex structures \cite{kaneda}. If so, large amplitudes of energy transfer are predominantly due to the stretching of such structures (not a new concept) rather than sequential grinding down of eddies, as has been depicted often.

A final comment seems to be in order on the implication of this work to transverse velocity increments, which are known to be symmetric. It is possible that the difference of scaling exponents between longitudinal and transverse exponents, highlighted several times in the literature \cite{DTS97,CSNC97,KIKRS20}, may be attributed to the difference between the negative increment $\delta_r u^-$ and the (symmetric) transverse increment, $\delta_r v$. This point needs to be explored further.

\section{Acknowledgments}
We thank P.K. Yeung and D.A. Donzis for their collaboration on numerical simulations. KRS thanks B.R. Dhruva and I. San Gil who explored these issues in their Ph.D. theses in the $1990$'s. This work used the Extreme Science and Engineering Discovery Environment (XSEDE), which is supported by National Science Foundation grant number ACI-1548562, on Stampede2 provided through Allocation TG-PHY200084. The computations were performed using supercomputing resources through the Blue Waters Project at the National Center for Supercomputing Applications at the University of Illinois (Urbana-Champaign).

\end{document}